\documentclass[
 reprint,
 amsmath,amssymb,
 aps,
]{revtex4-2}

\usepackage{xcolor}
\usepackage{graphicx}% Include figure files
\usepackage{dcolumn}% Align table columns on decimal point
\usepackage{bm}% bold math
\usepackage{amsmath}
\usepackage{booktabs}
\usepackage{tabu}
\usepackage{placeins}
\usepackage{braket}
\usepackage{hyperref}

\begin{document}

\preprint{APS/123-QED}

\title{Renormalization of States and Quasiparticles in Many-body Downfolding}

\author{Annabelle Canestraight}
\affiliation{Department of Chemical Engineering, University of California, Santa Barbara, CA 93106-9510, U.S.A.}
\email{acanestraight@ucsb.edu}
\author{Zhen Huang}
\affiliation{Department of Mathematics, University of California, Berkeley, CA 94720, U.S.A}
%\author{Lin Lin}
%\affiliation{Department of Mathematics, University of California, Berkeley, CA 94720, U.S.A}
%\affiliation{Applied Mathematics and Computational Research Division,
%Lawrence Berkeley National Laboratory, Berkeley, CA 94720, U.S.A}
\author{Vojtech Vlcek}
\affiliation{Department of Materials, University of California, Santa Barbara, CA 93106-9510, U.S.A.}
\affiliation{Department of Chemistry and Biochemistry, University of California, Santa Barbara, CA 93106-9510, U.S.A.}
%  \altaffiliation[Also at ]{Physics Department, XYZ University.}%Lines break automatically or can be forced with \\
% \author{Second Author}%
%  \email{Second.Author@institution.edu}
% \affiliation{%
%  Authors' institution and/or address\\
%  This line break forced with \textbackslash\textbackslash
% }%

% \collaboration{MUSO Collaboration}%\noaffiliation

% \author{Charlie Author}
%  \homepage{http://www.Second.institution.edu/~Charlie.Author}
% \affiliation{
%  Second institution and/or address\\
%  This line break forced% with \\
% }%
% \affiliation{
%  Third institution, the second for Charlie Author
% }%
% \author{Delta Author}
% \affiliation{%
%  Authors' institution and/or address\\
%  This line break forced with \textbackslash\textbackslash
% }%

% \collaboration{CLEO Collaboration}%\noaffiliation

% \date{\today}% It is always \today, today,
%              %  but any date may be explicitly specified

\begin{abstract}
We explore the principles of many-body Hamiltonian complexity reduction via downfolding on an effective low-dimensional representation. We present a unique measure of fidelity between the effective (reduced-rank) description and the full many-body treatment for arbitrary (i.e., ground and excited) states.  When the entire problem is mapped on a system of interacting quasiparticles [npj Computational Materials 9 (1), 126, 2023], the effective Hamiltonians can faithfully reproduce the physics only when a clear energy scale separation exists between the subsystems and its environment. We also demonstrate that it is necessary to include quasiparticle renormalization at distinct energy scales,  capturing the distinct interaction between subsystems and their surrounding environments. Numerical results from simple, exactly solvable models highlight the limitations and strengths of this approach, particularly for ground and low-lying excited states. This work lays the groundwork for applying dynamical downfolding techniques to problems concerned with (quantum) interfaces.
\end{abstract}

\maketitle

\section{Introduction}

The solution to a strongly correlated electronic structure problem necessitates a simplification. In many practical cases, such as for defects\cite{Romanova2023,Muchler2022Static} or molecules on surfaces\cite{Canestraight2024Efficient,rignanese2001quasiparticle,lei2023exceptional,egger2015reliable}, the key properties of interest are associated with only a selected subset of electronic states and excitations. In such cases, it is natural to develop an effective, or  ``downfolded,'' representation of the original problem in which the dimensionality is greatly reduced: the subsystem of interest is treated with a high-level theory, while the latter is approximated or treated using a mean-field or level of theory \cite{chang2023downfolding,Muchler2022Static,Pham2020-iu,Rusakov2019-wf,Sun2016-uu}. Such partitioning has a long tradition in quantum chemistry literature \cite{LowdinPartitioningI,LowdinPartitioningIV,SaadSpectralSchur}. The key question remains how to systematically develop such an effective Hamiltonian\cite{DvorakRinke2019(ED),Li2022-xr(ED),DvorakRinkeGolze2019ED,AuthierLoos2020,AFQMCEskridge2019}. 

In the condensed matter literature, it is typical to focus on the two-body interactions and their renormalization based on the Constrained Random Phase Approximation (cRPA) \cite{Aryasetiawan2009,MiyakeEffevtiveBand,GabiFreqDepndLoc,Chang2024,Bockstedte2018cRPA,Sheng2022cRPA,Ma2021EmbeddingcRPA}. Further, the renormalization is often taken in the static limit\cite{Muchler2022Static}, although a fully dynamical representation has been shown to significantly affect the results\cite{SakumaWerner2013Dynamical,Werner2010Dynamical,Werner2012LowEnergy}. Recently, alternative quantum chemistry approaches have been developed employing renormalization based on similarity transformation also yielding static Hamiltonians \cite{Bauman2019_2,Bauman2022,BaumanJCP2019,KowalskiFlow2019,shee2024static,Li2024-fg,He2020-ra}. An alternative, which we will explore here in more detail, is to approximate the problem by a set of interacting quasiparticles (QPs), i.e., renormalized electrons, within the subspace of interest. This was termed the dynamical downfolding in \cite{Romanova2023}. In this approach, we define the renormalization of the electron energies (and their interactions) through solving a complementary set of one and two-body propagator problems\cite{Romanova2023}. Such renormalizations can be computed efficiently within many-body perturbation theory even for large systems\cite{Vlcek2018Swift,Vlcek2017-rz}. The application to the negatively charged nitrogen-vacancy defect center in diamond \cite{Romanova2023} successfully reproduced its excitation energies known experimentally even with extremely small ``active space.''  Yet, it is not clear \textit{a priori} whether such a compression correctly captures the nature of the excitations (beyond their energy scales) and how to define the subsystem QPs for the cases when the particles can dissipate energy in the subsystem. 

Here, we provide a conceptual analysis of the downfolding procedure based on Hamiltonian compression and explore the QP couplings between the subsystem and the environment (the rest of the system). We illustrate these concepts on simplified, exactly solvable problems. Both analyses show that faithful downfolding on a set of interacting QPs requires a separation of energy scales between the subsystem and the environment and electron localization within the subsystem of interest. Both steps confirm that such an approach is suitable for e.g., quantum defects with excitations within a gap of the rest of the system. 

This paper is organized as follows: in Section \ref{sec:exactdownfolding}, we analyze exact many-body Hamiltonian downfolding using the Schur complement to establish a measure for the fidelity of this compressed representation. We demonstrate that in the context of many-body representation, the renormalization has a definite meaning relating the subspace solution to the full many-body wavefunction. Next, we analyze the use of single-particle Green's function (GF) to define a renormalized effective one-particle terms of the subspace Hamiltonian. In contrast to the original proposition\cite{Romanova2023}, we show that it is critical to include not only the single QP solutions, but also all ``satellites'' that capture the effective coupling to the rest of the system and hence provide further information about the renormalization by the environment.

\section{Exact Many-Body Downfolding: The Schur Complement}
\label{sec:exactdownfolding}

We assume that the system Hamiltonian is partitioned into a block structure consisting of a subspace of interest, represented by $\mathbf{H}_1$, and a rest space characterized by $\mathbf{H}_2$. These are coupled by off-diagonal blocks $\mathbf{C}$. The corresponding states are represented as tensor products of $\psi_j$ representing particular configurations in each subspace. In this context a particular choice of partitioning is such that $\mathbf{H}_1$ act on all configurations in the subspace of interest while the rest space is kept in a single selected configuration, e.g., in its ground state provided that the rest space excitations are energetically well separated. The particular choice of the partitioning determines the role of the downfolding and its physical interpretation discussed in next sections. The stationary states of the total system Hamiltonian and their energies thus constitute an eigenvalue problem:

\begin{equation}
    \label{eq.Schurproblem}
    \begin{bmatrix}
\mathbf{H}_1 & \mathbf{C} \\
\mathbf{C}^{\dagger} & \mathbf{H}_2\end{bmatrix} \begin{bmatrix}
\psi_1\\
\psi_2\end{bmatrix} = \varepsilon 
\begin{bmatrix}
\psi_1\\
\psi_2
\end{bmatrix}
\end{equation}
where $\varepsilon$ is the eigenstate energy.

In the context of downfolding, we seek to determine the energies of the stationary states by focusing only on a selected subspace of interest ( $\mathbf{H_1}$), which does not contain a portion of the eigenvector (in the rest space spanned by $\psi_2$). We thus introduce a renormalized (low dimensional) effective Hamiltonian employing on the Schur complement of the rest space
 \begin{equation}
 \label{Eq.Nleig}
     \mathbf{H}_{\rm eff}(\omega)=\mathbf{H}_1 + \mathbf{C} (\mathbf{I}\omega-\mathbf{H}_2)^{-1}\mathbf{C}^{\dagger} = \mathbf{H}_1 +  \mathbf{\Sigma}^{\rm S}(\omega),
 \end{equation}
where we introduce $\mathbf{\Sigma}^{\rm S}(\omega)$,  as the ``self-energy" that renormalizes terms in the $\mathbf{H}_1$ block. The reduced dimensionality translates Eq.~\ref{eq.Schurproblem} to a non-linear eigenvalue problem (in $\omega)$  acting only on $\psi_1$ space. Such a partitioning and renormalization has been explored in the past by Lowdin \cite{LowdinPartitioningIV,LowdinPartitioningI}, as well as in the context of the renormalization group \cite[Chap. 24]{Gustafson2020-xl}.

In principle, $\Sigma^S$
 has a pole structure determined by the rest space ($\mathbf{H_2}$). If the matrix $(\mathbf{I}\omega - \mathbf{H}_2)$ is invertible for all values of $\omega$, the non-linear eigenvalue problem $w=H_{\rm eff}(w)$
can be solved to find every eigenvalue of the full matrix in Eq. \ref{eq.Schurproblem} \cite{LowdinPartitioningI,martin_2016}.

This solution is demonstrated by the red point on the blue curve of Fig. \ref{fig1}. There is, however, not a unique way to generate an effective Hamiltonian with these eigenvalues, since it is based on an arbitrary choice of $\mathbf{H}_1$, however, physical interpretation of the self-energy is dependent on the basis.  Even if all the eigenvalues (or at least those of interest, e.g., the lowest lying ones) are obtained, the corresponding eigenvectors rely on the underlying choice of the subspace. For instance, when the matrix is partitioned into two subspaces, the eigenvectors of $\mathbf{H}_{\rm eff}$ naturally do not contain the entanglement between the spaces that may be seen in full eigenvector. In this scenario, the fixed point solution provides meaningful eigenvalues but the eigenvectors of $\mathbf{H}_{\rm eff}$ do not faithfully represent physical interpretation found in the eigenstates of $\mathbf{H}$. 

In the remainder of this section, we will demonstrate that the structure of self-energy, which renormalizes $\mathbf{H}_1$, is used to draw a connection between  the full many-body eigenstates and the eigenvectors of the effective (renormalized) Hamiltonian, i.e., it provides a direct access to the degree of separability between the subspaces. In Section \ref{sub:analyticalproof}, we will derive the mathematical relationship between the eigenstates of $\mathbf{H}$ and $\mathbf{H}_{\rm eff}(\omega)$. Following this derivation in Section \ref{sub:manybody}, we will demonstrate the relationship in a numerical proof, as well as discussing how this math can be interpreted physically in the case of a many-body Hamiltonian. Most importantly, we show that the relation between the exact eigenstates and the subspace eigenvectors is established merely from the knowledge of renormalized subspace, i.e., without the need to evaluate the full system Hamiltonian.

\subsection{Physical Interpretation of the Eigenvectors}
\label{sub:analyticalproof}

We will first derive the relation of the eigenvectors of the subspace and the stationary states of the full Hamiltonian. The eigenvectors of the Hamiltonian carry a critical information identifying the  particular types of excitations associated with each stationary state, characterized by its energy $\varepsilon$, in Eq.~\ref{eq.Schurproblem}. Therefore, it is essential to the physical interpretation of a downfolded Hamiltonian that this relationship is understood. Here, we will demonstrate that such information about the full eigenvectors can be constructed from the knowledge of $\mathbf{H}_{\rm eff}(\omega)$ alone.

We begin by considering the degree of renormalization of a given eigenvalue, which is associated with the magnitude of the self-energy at the fixed point solution. In analogy to the single QP picture \cite{Fetter} the typical measure is the ``renormalization factor'' representing the residue of the Schur complement.

\begin{align}
Z_i &= \left(1-\partial_\omega \left\langle \psi_i | \mathbf{H}_{\rm eff} (\omega) |\psi_i \right\rangle \middle |_{\omega=\varepsilon_i}\right)^{-1}\nonumber \\ &= \left(1-\partial_\omega \left\langle \psi_i | \mathbf{\Sigma}^{\rm S} (\omega) |\psi_i \right\rangle  |_{\omega=\varepsilon_i}\right)^{-1}.  
\end{align}

\begin{figure*}
    \centering
    \includegraphics[width=\columnwidth]{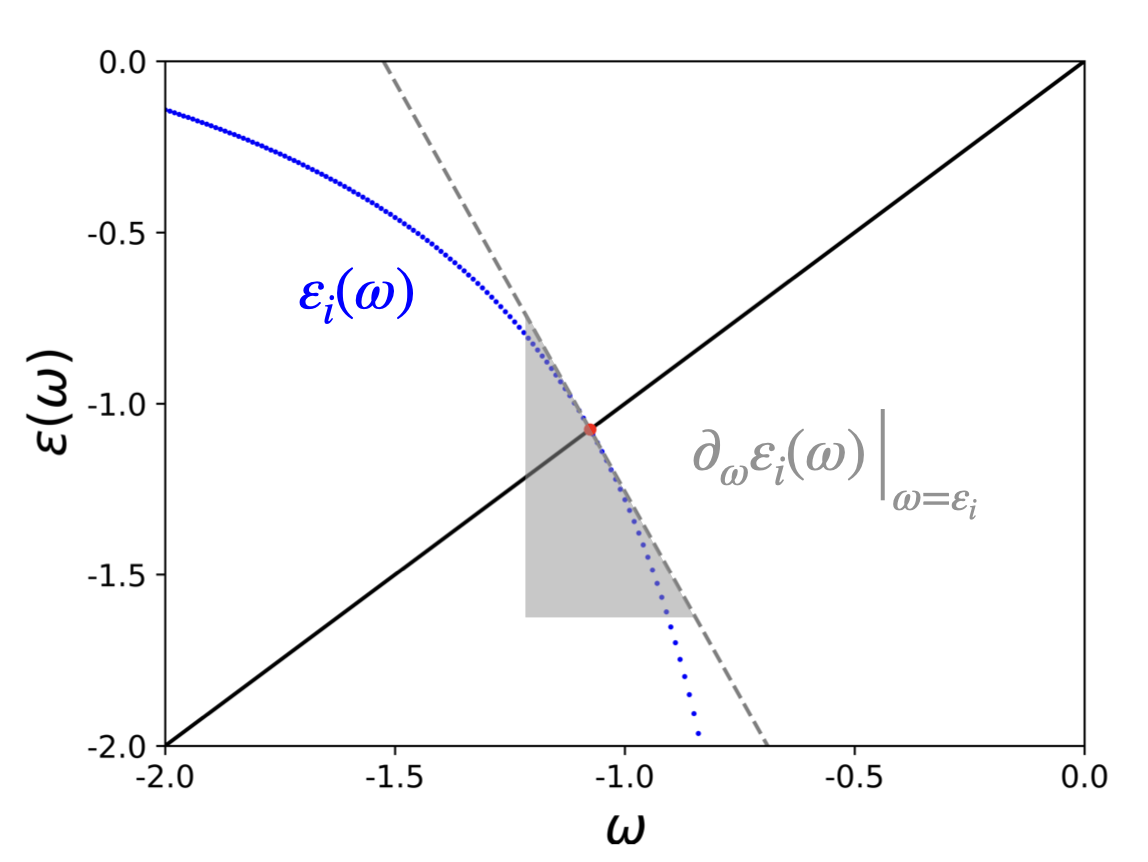}
    \caption{The fixed point solution for the eigenvalue is solved at the red point. The tangent line to this point gives the $Z$ factor.}
    \label{fig1}
\end{figure*}

Here, $Z_i$ is a scalar corresponding to the $i^{\rm th}$ eigenvalue of the Hamiltonian. The ``$Z$ Factor" is evaluated from the slope of the self-energy $\Sigma(\omega)$ when the frequency is equal to the eigenvalue $\varepsilon_i$ (demonstrated by the dashed, grey tangent-line in Fig. \ref{fig1}. 

The renormalization factor has been defined in the downfolding of a single-particle Hamiltonian, and we interpret it likewise for a many-body Hamiltonian. While it is well understood in the context of eigenvalue renormalization, we will now show how an analytical relationship between the eigenvector of the full Hamiltonian and the eigenvectors of  $\mathbf{H}_{\rm eff}$ is found to be given by $Z$. 

We consider a matrix of the following form: 
\begin{equation}
\mathbf{H}(\omega)=
\begin{bmatrix}
\mathbf{H}_1-\omega \mathbf I & \mathbf{C} \\ \mathbf{C}^{\dagger} & \mathbf{H}_2-\omega \mathbf{I}
\end{bmatrix} 
\end{equation}
Note that $H{(\varepsilon)}|\psi\rangle = 0$ is equivalent to the original many-body eigenvalue problem \eqref{eq.Schurproblem}.

The effective Hamiltonian $\mathbf H_{\text{eff}}(\omega)$ \eqref{Eq.Nleig} appears naturally if we perform the following matrix equivalence transformation of $H(\omega)$:

\begin{widetext}
\begin{equation}
\label{eq:mattrasnform}
\begin{split}
    \tilde{\mathbf{H}}=\begin{bmatrix}
        \mathbf{I}_1 & -\mathbf{C}(\mathbf{H}_2 - \omega \mathbf{I})^{-1} \\ 0 & \mathbf{I}_2
    \end{bmatrix} 
    \begin{bmatrix}
        \mathbf{H}_1-\omega \mathbf I & \mathbf{C} \\ \mathbf{C}^{\dagger} & \mathbf{H}_2 -\omega \mathbf{I}
    \end{bmatrix} 
    \begin{bmatrix}
        \mathbf{I}_1 & 0 \\ -(\mathbf{H}_2 -\omega \mathbf{I})^{-1} \mathbf{C}^{\dagger} & \mathbf{I}_2
    \end{bmatrix}  = \begin{bmatrix}
        \mathbf{H}_1-\omega \mathbf I-\mathbf{C}(\mathbf{H}_2-\omega \mathbf{I})^{-1} \mathbf{C}^{\dagger} & 0 \\ 0 & \mathbf{H}_2
    \end{bmatrix}
   \\ = \left[
    \begin{array}{cc}
     \mathbf H_{\text{eff}}(\omega)-\omega\mathbf{I}    & 0  \\
      0   & \mathbf H_2
    \end{array}
    \right]
\end{split}
\end{equation}
\end{widetext}

 We can see there is a self-energy, $\mathbf{\Sigma}(\omega)=\mathbf{C}(\mathbf{H}_2-\omega \mathbf{I})^{-1} \mathbf{C}^{\dagger}.$ 
 
Conceptually, the downfolding procedure encodes the information about the entire system into the subspace of $\mathbf{H}_1$ and the  problem is thus represented in two equivalent ways as:
\begin{align}
    \mathbf{H}\begin{bmatrix}
        \psi_1 \\ \psi_2
    \end{bmatrix} = 0 && \tilde{\mathbf{H}}\begin{bmatrix}
        \phi_1 \\ 0
    \end{bmatrix}=0
\end{align}
This is consistent with the fact that the eigenvectors of the effective Hamiltonian  $\tilde H$ span a subspace that is orthogonal to the $\psi_2$ components.

Given the form of the transformation in Eq. ~\ref{eq:mattrasnform}, the eigenvectors 
$\Psi_i =\begin{bmatrix}    \psi_1 \\ \psi_2
\end{bmatrix}_i$ and $\Phi_i=\begin{bmatrix}
    \phi_1 \\ 0
\end{bmatrix}_i$ are related by the following transformation\cite{SaadSpectralSchur}:
\begin{equation}
\begin{split}
    \begin{bmatrix}
    \psi_1 \\ \psi_2
\end{bmatrix}_i \propto \begin{bmatrix}
\mathbf{I}_1 & 0 \\ -(\mathbf{H}_2 -\varepsilon_i I)^{-1} \mathbf{C}^{\dagger} & \mathbf{I}_2    
\end{bmatrix} \begin{bmatrix}
    \phi_1 \\ 0
\end{bmatrix}_i \\  = \begin{bmatrix}  \phi_1 \\
      -(\mathbf{H}_2 -\varepsilon_i  \mathbf{I})^{-1} \mathbf{C}^{\dagger}\phi_1 
\end{bmatrix}_i
\end{split}
\end{equation}

$\mathbf{H}$ and $\mathbf{H}_{\rm eff}$ are both hermitian matrices with stationary eigenstates that must normalized to 1.
Consequently, we must normalize $\psi$:

\begin{align}
\label{eq:normalize}
\begin{split}
\begin{bmatrix}
    |\psi_1 \rangle \\ |\psi_2\rangle
\end{bmatrix}_i   = c_i
\begin{bmatrix}
    \mathbf{I} & 0 \\ 0 & -( \mathbf{H}_2 -\varepsilon_i  \mathbf{I})^{-1}  \mathbf{C}^{\dagger}
\end{bmatrix} 
\begin{bmatrix}
    \phi_1 \\ \phi_1
\end{bmatrix},\\
c_i = \frac{1}{\sqrt{1+\langle \phi_1|\textbf{C}(\textbf H_2-\varepsilon_i \textbf{I})^{-2}\textbf{C}^{\dagger}|\phi_1\rangle}},
\end{split}
\end{align}

Given the 2 sets of eigenvectors
$\begin{bmatrix}
    \psi_1 \\ \psi_2
\end{bmatrix}$, which belong to the full Hamiltonian and $\begin{bmatrix}
    \phi_1 \\ 0
\end{bmatrix}$, belonging to the effective Hamiltonian, we can measure their similarity through their inner product.  This yields the following relationship between the two sets of eigenvectors, where $i$ indexes over the individual eigenstates of both Hamiltonians. 

\begin{equation}
    \label{eq.endproof}
   Z_i = |\langle \Phi_i|\Psi_i \rangle|^{2} = |c_i|^2 = \frac{1}{1+\langle\phi_{1,i}|\mathbf{C}\left(\mathbf{H}_2-\varepsilon_i \mathbf{I}\right)^{-2} \mathbf{C}^{\dagger}| \phi_{1,i} \rangle}
\end{equation}

By inspection, we can see that the overlap of the eigenvectors is a function of the derivative of the  Schur self-energy with respect to $\omega$.
\begin{equation}
    \partial_\omega \langle \phi_{1,i}|\mathbf{\Sigma}^{\rm S}(\omega)  |\phi_{1,i}\rangle=\langle \phi_{1,i}| \mathbf{C} (\mathbf{H}_2 - \omega \mathbf{I} )^{-2} \mathbf{C}^{\dagger}|\phi_{1,i}\rangle
    \label{eq:derivative_self_energy}
\end{equation}
Note that  $|\phi_{1,i}\rangle$ also depends on $\omega$, and Eq. \ref{eq:derivative_self_energy} holds due to Hellman-Feynman theorem \cite{feynman1939forces}.
Substituting this into Eq. ~\ref{eq.endproof}, we see that this overlap is the same as the $Z$ factor, i.e.,  the slope of the self-energy at the fixed point solution. The difference between the true eigenvector of Eq. \ref{eq.Schurproblem} and the eigenvector of the effective Hamiltonian is thus given by the renormalization factor alone.

In fact, Eq.~\ref{eq:normalize} reveals that the unknown part of the full eigenvector is constructed directly from the knowledge of the spectral decomposition of $H_{\rm eff}$ (i.e., from $\phi_1$ and the self-energy). Indeed, the remaining component $ |\psi_{2}\rangle_i = \mathbf{C}^{-1} \mathbf{\Sigma}(\varepsilon_i)  |\phi_{1}\rangle_i  $ which requires the knowledge of only the Schur complement and the coupling block. One can use these to evaluate any expectation value  $\langle O\rangle_R = \left\langle \psi_2| \hat O | \psi_2\right\rangle $ for the rest-space. For large systems, this can not be done exactly, however, in cases where the action of the Schur complement and the coupling block are reliably approximated, these expectation values can be estimated. We do not utilize this reconstruction further here, although future work will explore this relation further.

\begin{figure*}[h]
    \centering
    \includegraphics[width=\textwidth]{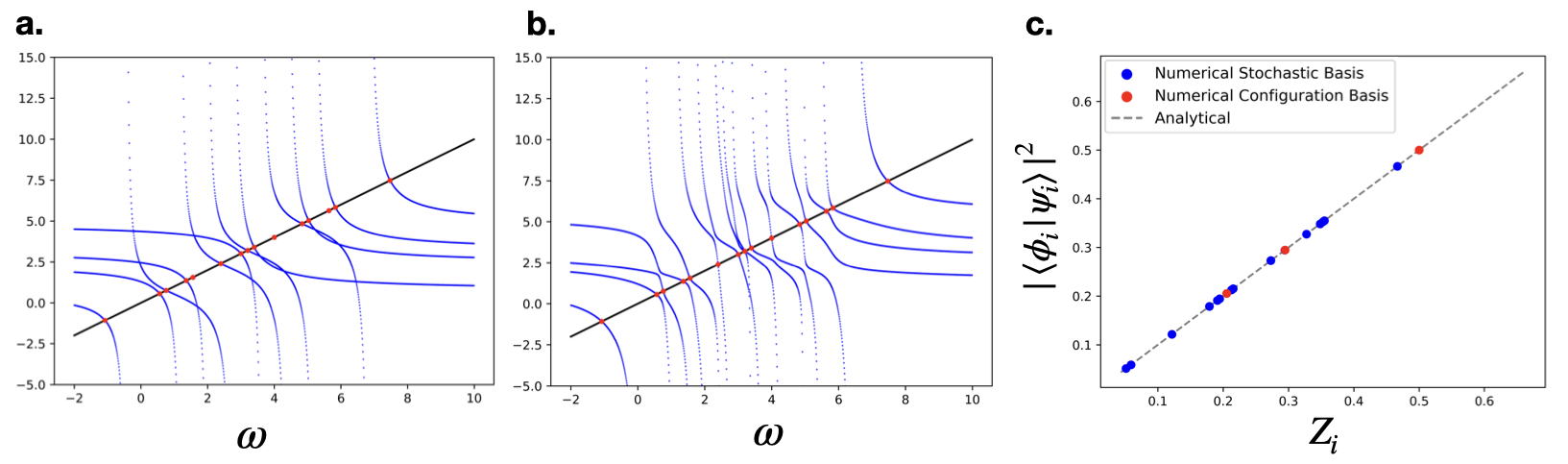}
    
    \caption{a. We show a fixed point solution where some ED solutions are missing due to the orthogonality of the eigenvectors to the downfolded subspace. In b. we demonstrate how a random unitary transformation of the Hamiltonian assures that a fixed point solution is found for every ED solution. In c. we use the fixed point solution in plots a and b to give numerical proof of Eq. \ref{eq.endproof} }
    \label{fig:2}
\end{figure*}

As mentioned earlier, the choice of partitioning can generally be arbitrary. For some choices of the subspace Hamiltonian, however, the problem is not invertible (as the particular $\varepsilon$ is an eigenvalue of $\mathbf{H}_2$). $\mathbf{C}$ is then sparse and does couple all excitations in the rest space to $\mathbf{H}_1$. In other words, the $i^{\rm th}$ eigenvector of the full Hamiltonian is entirely orthogonal to the states of the downfolded subspace and $Z_i=0$. As a result, there are fewer poles in the self-energy $\Sigma^{\rm S}(\omega)$, and the fixed point solution to $\mathbf{H}_{\rm eff}(\omega)$ will miss such eigenvalues of the original problem\cite{LowdinPartitioningI}. In the rest of this paper, we will assume that all eigenvalues are accessible, in principle. This is accomplished, for instance, by employing a random congruent transform of the basis, as shown in the next subsection. Through this unitary transformation all eigenvalues are preserved, and it is unlikely that any eigenvector is orthogonal to the downfolded subspace. In practice this yields $Z_i>0, \quad \forall i$.

\subsection{Numerical example and Physical Interpretation of Exact Downfolding}
\label{sub:manybody}

To this point, the mathematics and interpretation has been general to any Hamiltonian or matrix. The choice of downfolding, however, leads to unique interpretations of a many-body problem. In this subsection, we will analyze the particular consequences of the exact downfolding procedure for the physical interpretation of interactions in a many-body system. Without loss of generality, we consider a Hamiltonian of the following form:
\begin{equation}
 \label{extendedhubbard}
     \hat{\mathbf{H}}=-\sum_{i,j,\sigma}t_{ij}\hat{c}_{i\sigma}^{\dagger}\hat{c}_{j\sigma} +\frac{1}{2}\sum_{i,j,\sigma,\sigma'}U_{ij}\hat{c}_{i\sigma'}^{\dagger}\hat{c}_{i\sigma'}\hat{c}_{j \sigma}^{\dagger}\hat{c}_{j\sigma}
 \end{equation}

Here, $t_{ij}$ represents the one-body terms (both on-site and inter-site, or ``hopping'' amplitudes), and $U_{ij}$ are the two-body (long-range) density-density interaction terms. This Hamiltonian acts on a basis of configuration states, which can be represented as product states of a subsystem of interest and its environment.

In Fig. ~\ref{fig:2}, we demonstrate the fixed point solution for a example Hamiltonian in multiple bases: one of many-body configurations and one stochastic basis, where each state is a random linear combination of configuration states.  In Fig.~\ref{fig:2}~a, we show the fixed point solution in the configuration basis, where some ED eigenvalues (red points) do not correspond to a fixed point solution and are lost by downfolding. This means that the excitation is in the orthogonal subspace. We then use a random congruent transform to change the basis of this matrix, and we partition it again into subspace and rest-space blocks. In  Fig~\ref{fig:2}~b more poles are introduced to $\mathbf{\Sigma}^{\rm S}(\omega)$ and all of eigenvalues can be found by fixed point solution. Using these 2 sets, we give a numerical proof of the $Z$ factor in Eq. \ref{eq.endproof} by computing both the slope and the overlap of the eigenvectors. In Fig.~\ref{fig:2}~c it is shown that in both bases, all solutions follow the established relationship, although they arise from different partitioning.

Although we have shown that the choice of basis can guarantee all eigenvalues of the original Hamiltonian can be found, we will now choose to work in a basis where each element of $\mathbf{\Sigma}^{\rm S}$ can be clearly interpreted as the renormalization of an individual term of the extended Hubbard Hamiltonian. In this basis, some excitations confined to the $\psi_2$ subspace are missing, however, we chose our subspace such that the select solutions are guaranteed to be present. 

The physical interpretation of the excitations depends strongly on the form of the the effective Hamiltonian (i.e.,  the choice of the downfolding subspace and basis). For instance, even when the full Hamiltonian matrix is sparse and contains only diagonal density-density interactions and single particle hopping terms, the its compressed version is does not preserve the original sparsity as the Schur complement constructed from this matrix is dense. The renormalized Hamiltonian contains entries that can be interpreted as two-body interactions beyond density-density terms.

Similarly ambiguity arises in the diagonal terms of the matrix where, in a basis of configuration states, the effect of the Schur complement can be equivalently thought of as renormalization of the one-body on-site potential or of the density-density Hartree term. Therefore, in the basis of configurations, we can (arbitrarily) define $\mathbf{\Sigma^{\rm S}}$ to renormalize each of the individual one-body and two-body interactions in $\mathbf{H}_1$.

In practice, we aim for a partitioning defined either energetically (e.g., on the low energy sector) or spatially (on selected sites). 

While the subspace is spanned by all configurations of the subsystem, the environment is held in a fixed configuration (or a selected small number of them). The rest-space contains (in principle) all of the remaining configurations of the environment. Generally, downfolding is performed in a basis where that spans multiple configurations of the subspace of interest, but fixes the environment to be in its ground state \cite{Romanova2023,DvorakRinke2019(ED),He2020-ra,Bauman2019_2} In this work, we will downfold on such states of the form 
\begin{equation}
   |\Psi^{N}\rangle = |S^N_i\rangle \otimes |R_0^N\rangle 
\end{equation}
where both subsystems have a conserved number of particles, and the environment, $R$, is in its ground state, while the subsystem of interest, $S$ can span all configurations.

%The Schur complement  yields renormalized  one and two-body terms, $\tilde{t}$ and $\tilde{U}$. 
Downfolding is typically thought of as a renormalization of one and/or two-body terms by the environment \cite{Aryasetiawan2009,MiyakeEffevtiveBand,GabiFreqDepndLoc,Chang2024,Bockstedte2018cRPA,Sheng2022cRPA,Ma2021EmbeddingcRPA,Romanova2023,Muchler2022Static,SakumaWerner2013Dynamical,Werner2010Dynamical,Werner2012LowEnergy}. When the downfolding subspace is in a configuration basis, the Schur complement gives the renormalization of individual one and two-body terms, and the effective Hamiltonian takes the following form:
\begin{equation}
    \label{eq.Htilde}
     \tilde{\mathbf{H}}_1=-\sum_{i,j,\sigma}\tilde{t}_{ij}\hat{c}_{i\sigma}^{\dagger}\hat{c}_{j\sigma} +\frac{1}{2}\sum_{i,j,k,l,\sigma,\sigma'}\tilde{U}_{ijkl}\hat{c}_{i\sigma'}^{\dagger}\hat{c}_{j\sigma'}\hat{c}_{k \sigma}^{\dagger}\hat{c}_{l\sigma}
\end{equation}

Here, $\tilde{t}_{ij}$ of the effective Hamiltonian contains an additional self-energy correction ($\Sigma^{\rm env}_{ij}$)from the rest-space on which $\tilde{\mathbf{H}}$ does not act explicitly.
\begin{equation}
\label{eq:sprenorm}
    \tilde{t}_{ij} = t_{ij} + \Sigma^{\rm env}_{ij}
\end{equation}

Further, in this basis, where the matrix has been partitioned such that the downfolding subspace contains only variations of the subsystem of interest, the many body renormalization factor $Z$ gives a measure of how well an excitation is contained within that subsystem. In many approximate downfolding methods, a large Hamiltonian is mapped onto a single static matrix of a smaller dimension. The eigenstates correspond to the true many-body states when $Z \approx 1$, meaning the excitation is well contained in the downfolding subspace. For a system where the subsystem and environment are energetically degenerate, such clear separation obviously fails. In figure \ref{fig:3}, we numerically show this to be true by performing the exact downfolding on a model system where the energy separation between the subsystem and environment is varied:

\begin{figure*}[h]
    \centering
    \includegraphics[width=\textwidth]{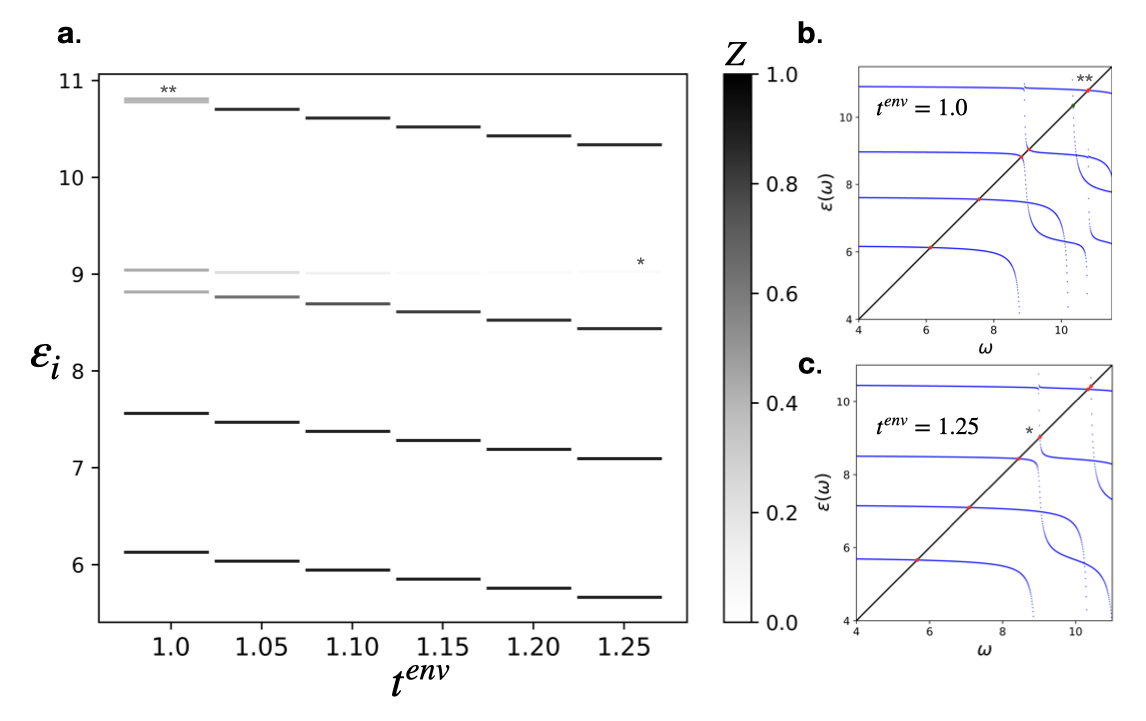}
    \caption{a. A portion of the eigenvalue spectrum is given, with the color bar indicating the $Z$ factor associated with each eigenvalue. As the energy scale of the environment is separated from that of the subsystem, the $Z$ factor of the ``intruder" states decreases. In part b.  for $t^{env}=1$ the solutions occur at a pole and have smaller $Z$ factors, while for part c. $t^{env}=1.25$ and the $Z$ factors have split.   }
    \label{fig:3}
\end{figure*}
In Fig. ~\ref{fig:3} the color of each spectral line gives the $Z$ factor for each of the eigenstates. In Fig ~\ref{fig:3}~a , we see that for no energy separation ($t^{env}=1$) that rather that there are $>4$ eigenvalues and some are split between 2 values with nearly equal $Z$ factor. As $t^{env}$ is increased and we move across the plot, these split values move apart and one takes on a larger $Z$ factor while the other disappears as an ``intruder state" that have strong fluctuations in the environment. For $t^{env}=1$ we can clearly see 4 eigenvalues that correspond to the 4 eigenvalues of the subspace Hamiltonian $\mathbf{H}_1$. This effect can be interpreted in Fig.~\ref{fig:3}~b where the fixed point solution is shown for $t^{env}=1$. Here we see that the solution stemming from the 3rd and 4th eigenvalues of $\mathbf{H}_1$ occur at poles of the self-energy, meaning they occur near an excitation of the environment. In Fig. ~\ref{fig:3}~c, when $t^{env}=1.25$, we see that the solution is now to the left of the pole and the 2 solutions at each pole have very different $Z$ factors. In other words, it becomes possible to distinguish the excitations as occurring in the subsystem of interest versus the environment subsystem. We conclude that, for an approximately renormalized Hamiltonian $\tilde{\mathbf{H}}$ to have eigenstates that correspond to ED eigenstates of the original Hamiltonian, a separation of energy scales is required between the subsystem and environment. 

While downfolding via the Schur complement is intractable for realistic systems, as its construction is equivalent to solving the full many-body problem, the results shown above demonstrate that a near product-state form of the eigenvectors, which arises from energy scale separation, is necessary for physically meaningful eigenstates to be found from a static renormalized effective Hamiltonian. In the following section, we will discuss how such an effective Hamiltonian can be formulated using Green's functions, and how the Green's function can inform us of when the true eigenvectors are expected to take a product-state form.

\section{Dynamical Downfolding: Renormalization in a system of quasiparticles}
\label{sec:1body}

In this section, we shift our focus from the explicit construction of the self-energy via the Schur complement to the Dynamical Downfolding method. This approach leverages Green’s functions to solve for individual terms of the many-body Hamiltonian. We assume that the environment ``renormalizes" the subsystem, effectively describing it as a set of interacting quasiparticles~\cite{Romanova2023}. These individual quasiparticles/quasiholes are defined in the context of the one-body Green's Function, which utilizes the same conceptual machinery as the many-body downfolding but on the (effective) single-particle basis. The goal of the dynamical renormalization is to construct an effective static Hamiltonian that represents the ground and some of the excited states of the subsystem. We will now discuss the intricacies of the Hamiltonian construction using the reduced information from the one-particle Green’s function. Unlike in the Schur complement construction focusing on each non-linear solution separately, we show that the renormalization requires using the entire (multipole) structure of the one-QP Green's function.

In this section, we first comment on the origin of multiple poles and follow with the analysis of their contribution to the renormalization of QPs 

\subsection{1-Body Green's Function}
  \label{sub:1-bodyGF}
We relate the renormalization of one-body terms in $\mathbf{H}_1$ to the effective one-body Hamiltonian $H^{QP}$, which is characterized by its resolvent, the single-particle GF 
\begin{equation}
G_{ij}(t-t') = \langle \Psi_0^{N} | \mathcal{T} [\hat{c}_j(t'), \hat{c}_i^{\dagger}(t)] |\Psi_0^{N} \rangle
\end{equation}
Here, $\hat{c}^{\dagger}_j$ and $\hat{c}_i$ are creation/ annhilation operators on the sites of the subspace ($i$ and $j$), and $\mathcal{T}$ is the time-ordering operator; the  $\Psi_0^{N}$ is the $N$ particle ground state of the system. The GF represents the probability amplitude associated with adding/removing a particle at with time delay $t-t'$ (at equilibrium). This quantity is determined by an effective dynamics of a single (quasi-)particle entering the the one body portion of the Hamiltonian in Eq. \ref{extendedhubbard}.

The Lehmann representation of the GF provides a convenient form of the equilibrium GF:
\begin{equation}
\label{eq:lehmann}
\begin{split}
     G(\omega)_{ij}=\lim_{ \eta \to 0^+}\sum_{k} \frac{\langle \Psi_0^{N}|\hat{c}_j^{\dagger}|\Psi_k^{N-1}\rangle\langle\Psi^{N-1}_k|\hat{c}_i|\Psi_0^{N}\rangle}{\omega-\tilde{\varepsilon}_k^h +i\eta}
     + \\  \sum_{k} \frac{\langle \Psi_0^{N}|\hat{c}_j|\Psi_k^{N+1}\rangle\langle\Psi^{N+1}_k|\hat{c}^{\dagger}_i|\Psi_0^{N}\rangle}{\omega-\tilde{\varepsilon}^{p}_k -i\eta}
    \end{split}
\end{equation}
 $\mathbf{G}(\omega)$  is a complex valued function for which the real part has poles located at the single QP energies, $\tilde{\varepsilon}_k$, eigenvalues of an effective (renormalized) single-particle Hamiltonian $\hat{H}^{QP}(\omega)$. The imaginary part of $\mathbf{G}(\omega)$ has peaks located at the QP energies. The fixed point solutions to the QP Equation,
\begin{equation}
\label{eq:QPeq}
    \tilde{\varepsilon}_k |D_k\rangle=\hat{H}^{QP}(\tilde{\varepsilon}_k)|D_k\rangle,
\end{equation}
yields the QP energies and the Dyson orbitals, $|D\rangle$. In the following, we represent the single particle position states as a linear combination of the Dyson orbitals \footnote{We note that Dyson orbitals naturally form a non-orthogonal basis owing to the frequency dependence of the quasiparticle Hamiltonian}:
 \begin{equation}
    \label{positiondyson}
        |w_i\rangle =\sum_k \alpha_k |D_k\rangle 
 \end{equation}  
Here, $|w_i\rangle$ is a position state of a particle on site $i$ and $\alpha_k=\langle w | D_k\rangle$. The $i,j$ elements of the Green's function differ only by the relative amplitudes of the peaks of the imaginary part of the GF (i.e., the values $\alpha_k$), determined by the projection of the Dyson orbitals onto the position states. 

Using these Dyson orbitals, energy eigenvectors of the single-particle Hamiltonian, we define the single-particle (onsite) energy in a position basis \cite{Manne1970}: 
\begin{equation}
\label{eq. weighted average}
    \varepsilon_{ij} = \sum_k \alpha_{ik} \alpha^*_{jk} \tilde{\varepsilon}_k
\end{equation}
Spectral weight transfer to the satellites results in a shift in the single-particle energy, which is a weighted average over the hole peaks of the GF.

In practice, we use the one-body GF to compute the energy of a single particle in the subsystem of interest with and without environment response. This is done by defining the $N \pm 1$ particle Hilbert space with and without multiple configurations of the environment (see SI for details). The response of the environment introduces additional poles to the GF, which account for the coupling of the particle to neutral excitations of the environment. The difference of the two single-particle energies we obtain is the one-body environment self-energy $\Sigma^{\rm env}$, and it is incorporated into the effective Hamiltonian as shown in Eq. \ref{eq:sprenorm}. %\footnote{When there are no interactions between subsystems, a particle created or annihilated in the subsystem of interest does not affect the environment, which remains in its ground state. Consequently, the two GF's we compute are the identical and $\Sigma^{\rm env}_{ij} = 0 \, \forall \, {i,j}$}.

\subsection{Renormalization Factor in the 1-Body Green's Function}\label{ssec:3b-Zfactor1GF}

The many-body renormalization factor $Z$ can not be computed exactly from the Green's function, however, information about the structure of the true eigenvectors may be found from the satellites of the GF. We will now consider how excitations of the environment correspond to spectral weight transfer to the higher energy satellites, indicating entanglement between the subsystems.

For the description of the many-body states, we consider a product form basis:
   \begin{equation}
    \label{Eq.npart1}
        |\Psi^N\rangle = |S^N\rangle \otimes |R^N\rangle
    \end{equation}
Here, $S^N$ denotes an $N$ particle state of the subsystem and $R$ is a state of the environment. We consider the case of no hopping between the subsystem and the environment\footnote{When there are no interactions between subsystems, a particle created or annihilated in the subsystem of interest does not affect the environment, which remains in its ground state. Consequently, the two GFs we compute are identical and $\Sigma^{\rm env}_{ij} = 0 \, \forall \, {i,j}$}. When we compute the full one-body Green's function, the peaks represent states following general form
   $ |\Psi^{N \pm 1}\rangle = |S^{N \pm 1}\rangle \otimes |R'^N\rangle $

Here, a particle which is added or removed from the subsystem $S$ can couple to neutral excitation of $S$ and or $R$. The primary QP peak of the GF corresponds to no excitation of the system.

When the environment is fixed in its ground state as an external potential, however, all satellite peaks of the GF represent excitations of $S$ and $R=R'$. Therefore, peaks which are exclusive to the case of a general $R'$ indicate that the particle/hole is coupled to some excitation of the environment.

When there is a substantial separation of the energy scales of the subsystem of interest and environment, the additional peaks in the $R'$ case will be energetically distinct from those in the $R=R'$ case. This indicates that many-body excitations of the subsystem are unentangled and the many-body renormalization factor $Z\approx 1$ and the resulting eigenvectors can be high-fidelity.

\section{Numerical Results}
\label{sec:results}

We now illustrate the Dynamical Downfolding method and show that this method generally succeeds for the ground states but gradually fails for higher energy excited states. These calculations were performed for multiple values of $U$ on a particular example system selected from the Exact Downfolding: two dimers at half-filling placed end to end (See SI for details). We downfold one dimer onto the other.

  \begin{figure*}[h]
    \centering
    \includegraphics[width=\textwidth]{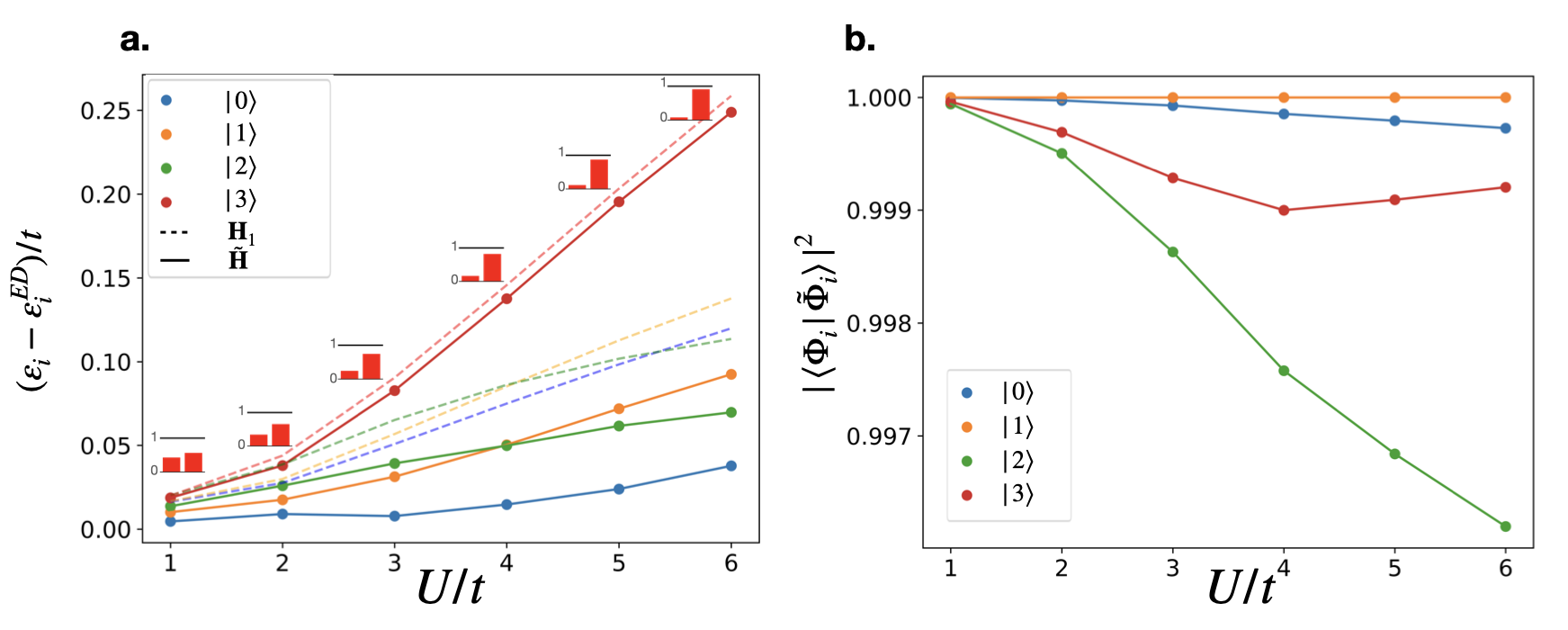}
    \caption{\textbf{a.} For each of the 4 states, across 6 values of U, we have plotted the distance between the approximated eigenvalues and the ED eigenvalues. The solid lines/points show the 1-body renormalized result and the dashed lines show the PSA with no renormalization. The bar plots show the particle density on each site of the subsystem dimer for the highest excited state. \textbf{b.} We take the overlap of the exact-downfolding eigenvectors and the $\tilde{H}$ eigenvectors for each value of U. }
    \label{fig:4}
\end{figure*}

In Fig. \ref{fig:4} a. we see that for the lowest 3 eigenvalues (corresponding to states $|0\rangle\,,|1\rangle\,,|2\rangle$), there is a large improvement when the 1-body renormalization is applied for all 6 values of $U$. This is most notable in the ground state $|0\rangle$, which, without renormalization has very similar error to $|1\rangle$ and $|2\rangle$ excited states but is much lower after renormalization. The highest excited state $|3\rangle$ is unaffected by the renormalization and shows the largest error.

Next, we will consider whether $\tilde{H}$ yields eigenvalues only, or whether the eigenvectors also approximate the ED eigenvectors. In Fig. ~\ref{fig:4}~b we show the overlaps of the $\tilde{H}$ eigenvectors $\tilde{\Phi}$ with the $\mathbf{H}_{\rm eff}(\omega)$ eigenvectors $\Phi$. For all eigenstates, we see that the overlap is close to unity. Eigenvectors $|0\rangle$ and $|1\rangle$ are highly accurate \footnote{In the case of the $|1\rangle$ eigenvector, this state is always a product state and is always the same across all values of $U$.} For $|2\rangle$ and $|3\rangle$, we see that the overlap decreases more significantly as $U$ increases. We note that there is not direct correlation in eigenvalue and eigenvector accuracy. $|3\rangle$ is a higher fidelity eigenvector than $|2\rangle$ , although the corresponding eigenvalue is much worse. 

We will now comment on the limitations of Dynamical Downfolding based on the physical interpretations of the third excited state $|3\rangle$. The bar-plot insets to Fig. ~\ref{fig:4}a~, show the particle density on the left and right sites of the subsystem dimer for $|3\rangle$. As $U$ increases, density shifts from being nearly equal across the two sights, to being heavily localized on the right-hand side, which is in closed proximity to the environment dimer. The energy of this state is heavily determined by the two-body interactions between the subspace particles as well with the environment. Such a physical analysis of the eigenstate gives us a means of predicting which eigenvalues of $\tilde{H}$ are accurate without knowing the ED spectrum.

In practice, we observe that the accuracy of each eigenvalue is correlated with the sensitivity of the solutions to the variations in the environment self-energy. The least accurate eigenvalues are insensitive to a change in the one-body self-energy. We also note that at this point, there is no obvious and simple route to define a renormalized two-body interaction that would satisfactorily reproduce the spectrum; in principle, such a solution needs to be obtained by solving the effective two-body propagator for all pairs of quasiparticles. This is, in turn, a multiparameter non-linear eigenvalue problem, which needs to be further approximated as discussed in\cite{Romanova2023}.

\section{Outlook and Discussion}

In conclusion, we have analyzed multiple downfolding approaches and illustrated them on a small, exactly solvable model system. Specifically, we explored an approach using the many-body Schur complement of the full Hamiltonian and an approach based on the map on effective renormalized quasiparticles (i.e., invoking the single-particle renormalization scheme). In Section \ref{sec:exactdownfolding}, we demonstrate that for any matrix compressed via its Schur complement, the renormalization factor is fully defined by the overlap between the true (ED) eigenvector and the downfolded eigenvector. Thus, the renormalization factor serves as a fidelity metric for the downfolded eigenvectors. Not surprisingly, the  downfolded Hamiltonian represents the original problem (i.e., the portion of the energy spectrum) if there is a sufficient energy separation between the subsystem of interest and the environment. This indicates that the excitations represented by the eigenvectors are contained in the correlated subsystem. In this case, the true stationary states are well captured by a product state between the environment and the subsystems. 

In Section \ref{sec:1body}a, we demonstrate how an effective many-body Hamiltonian can be constructed from a subspace of interacting quasiparticles derived from one-body Green's Functions (GF). This Dynamical Downfolding approach was heuristically suggested in Ref.~\cite{Romanova2023}, while we provide a more in-depth discussion of its formulation here. In this first-order approximation, all single-particle terms of the Hamiltonian are corrected by the single-particle environment self-energy. Notably, the exact GF yields multiple solutions to the quasiparticle equations, accounting for the correlation between the inserted particle/hole and a neutral excitation of the system. For the first time, we provide a method for including these multiple solutions in the renormalization of the single-particle energy based on their projection onto the single-particle states in a chosen basis (i.e., position states). Additionally, in Section \ref{sec:1body} b, we discuss how the position and height of the GF’s satellite peaks can serve as diagnostics for the relative energy scales of subsystems, indicating the reliability of the downfolded eigenvectors. This is a key to interpreting the results of downfolding without finding the many-body renormalization factor $Z$.

In Section \ref{sec:results}, we apply Dynamical Downfolding of one-body terms to a fully solvable model system. By varying the Hubbard interaction parameter $U$, we observe that incorporating satellite solutions consistently and \textit{significantly} reduces the magnitude of the one-body environment self-energy. Notably, for our largest value, $U = 6$, the single-particle environment self-energy is reduced by up to $42\%$. This finding highlights the need to approximate the self-energy using methods beyond $GW$—such as $GW \Gamma$—which are known to provide multiple solutions to the QP equation (e.g., due to coupling to spin fluctuations, etc.)\cite{CarlosGWGamma,Weng2023vertex,KutapovVertex,Maggio2017-bm,KresseVertex}. Additional methods for improving the satellite solutions from the non-interacting GF have been pioneered \cite{Cristostomo2024,El-Sahili2024}. Note that accurate downfolded representations do not necessitate that spectral weight transfer carries a well-defined physical meaning. Indeed, a fictitious coupling to charge neutral (bosonic) excitations at the ``average'' energy of the environment excitations would suffice to reproduce the energetics; this approach would connect the dynamical downfolding with the recent advances in Gutzwiller and slave boson approaches\cite{CarlosGGA,LanataGGA,PhysRevX.11.041040,GabiRISB2007}. 

As shown (heuristically) before, Dynamical Downfolding approaches can be (and have been) readily applied to realistic systems using many-body perturbation theory (e.g., with $GW$ approximation)\cite{Romanova2023}. This work provides a jumping-off point for the future development of the renormalized compression techniques that leverage Green's function framework with well-defined rules for ensuring the fidelity of the reduced problem solutions.

\section*{Acknowledgements}
  The theoretical development and numerical implementation (A.C. and V.V.), and the Analytical proof (Z.H.) are based upon work supported by the U.S. Department of Energy, Office of Science, Office of Advanced Scientific Computing Research and Office of Basic Energy Sciences, Scientific Discovery through Advanced Computing (SciDAC) program under Award Number DE-SC0022198. The numerical results fo the Green's function downfolding calculations performed by A.C. were supported by Wellcome Leap as part of the Quantum for Bio Program. Finally, the authors would like to thank Carlos Mejuto-Zaera, Lin Lin, and Cian Reeves for many insightful discussions.

\medskip

\bibliography{bib}% Produces the bibliography via BibTeX.

\end{document}